\documentclass[aps,prl,twocolumn,showfootnotes,showpacs]{revtex4} 

\usepackage{epstopdf}
\usepackage{graphicx}

\input epsf

\def\lsim{\mathrel{\raise.4ex\hbox{$<$\kern-.75em\lower1ex\hbox{$\sim$}}}}
\def\gsim{\mathrel{\raise.4ex\hbox{$>$\kern-.75em\lower1ex\hbox{$\sim$}}}}

\def\cmm2{{\,\rm cm^{-2}}}
\def\cm2{{\,{\rm cm}^2}}
\def\cmm3{{\,{\rm cm}^{-3}}}
\def\gcmm3{{\,{\rm g\,cm^{-3}}}}

\def\fun#1#2{\lower3.6pt\vbox{\baselineskip0pt\lineskip.9pt
  \ialign{$\mathsurround=0pt#1\hfil##\hfil$\crcr#2\crcr\sim\crcr}}}

\def\be{\begin{equation}}
\def\ee{\end{equation}}
\def\bea{\begin{eqnarray}}
\def\eea{\end{eqnarray}}

\begin{document}

\title{Extended MSSM Neutralinos as the Source of the PAMELA Positron Excess}

\author{Dan Hooper$^{1,2}$ and Tim M.P. Tait$^{3.4}$}
\affiliation{$^1$Center for Particle Astrophysics, Fermi National
Accelerator Laboratory}
\affiliation{$^2$Department of Astronomy \& Astrophysics, The
University of Chicago}
\affiliation{$^3$Department of Physics \& Astronomy, Northwestern University}
\affiliation{$^4$HEP Division, Argonne National Laboratory}
\date{\today}

\begin{abstract}
We consider a scenario within the Minimal Supersymmetric Standard Model extended by a singlet chiral
superfield, in which neutralino dark matter annihilates to light singlet-like Higgs bosons, which proceed to decay to either electron-positron or muon-antimuon pairs. Unlike neutralino annihilations in the MSSM, this model can provide a good fit to the PAMELA cosmic ray positron fraction excess. Furthermore, the singlet-like scalar Higgs can induce a large Sommerfeld enhancement and provide an annihilation rate sufficient to accommodate the observed positron excess.
\end{abstract}
\pacs{95.35.+d; 95.85.Ry; 11.30.Pb; 14.80.Ly; ANL-HEP-PR-09-36; FERMILAB-PUB-09-258-A; NUHEP-TH/09-07}
\maketitle

The satellite-based experiment PAMELA has reported a cosmic ray positron fraction (defined as the ratio of positrons to electrons-plus-positrons) which rises rapidly between 10 GeV and 100 GeV~\cite{pamela}. This is in stark contract to the behavior predicted for positrons produced through interactions of cosmic ray protons with the interstellar medium~\cite{secbg}. Although the origin of this positron excess is currently unknown, a number of plausible sources have been proposed, including pulsars~\cite{pulsars}, the acceleration of positron secondaries in cosmic ray acceleration regions~\cite{secacc}, and dark matter annihilations~\cite{annihilations1,annihilations2,lepsom,leptons,antiprotons,antiprotons2,gammarayssyn,sommerfeld,sommerfeld2} or decays~\cite{decay}.

Efforts to explain these observations with annihilating dark matter face some challenges. In particular, the rapid rise of the PAMELA positron fraction appears to require a very hard injected spectrum, which in turn requires the responsible dark matter particle to annihilate primarily to charged leptons~\cite{annihilations1,annihilations2,lepsom,leptons} (for an exception, see Ref.~\cite{clump}). Furthermore, dark matter candidates which annihilate largely to quarks or gauge bosons are also predicted to overproduce cosmic ray antiprotons~\cite{antiprotons,antiprotons2}, gamma rays, and synchrotron emission~\cite{gammarayssyn} if the overall annihilation rate is normalized to produce the positron fraction reported by PAMELA. 
Within the context of the Minimal Supersymmetric Standard Model (MSSM), neutralino dark matter annihilates largely to final states consisting of heavy quarks or gauge and/or Higgs bosons~\cite{jungman}. As a result, such annihilations produce a relatively soft spectrum of cosmic ray positrons~\cite{neu} and are unable to provide a viable explanation for the PAMELA excess. 

The same conclusion is not necessarily reached in supersymmetric models with an extended Higgs sector. 
Extensions of the MSSM by a singlet chiral superfield are motivated in order to explain the size of
the $\mu$ term \cite{nmssm}, to raise the mass of the lightest CP even Higgs boson above the LEP II bound
\cite{BasteroGil:2000bw}, to reduce electroweak fine-tuning \cite{Dermisek:2005ar}, or to catalyze electroweak baryogenesis \cite{Funakubo:2002yb}.  Such extensions are described by superpotential,
\begin{eqnarray}
v_0^2 \hat{S} + \frac{1}{2} \mu_S \hat{S}^2 + \mu \hat{H}_u \hat{H}_d + \lambda \hat{S} \hat{H}_u \hat{H}_d
+ \frac{1}{3} \kappa \hat{S}^3 ~,
\label{eq:W}
\end{eqnarray}
and soft Lagrangian,
\begin{eqnarray}
\frac{1}{2} m_S^2 |S|^2 + B_S S^2+ \lambda A_\lambda S H_u H_d +  \kappa A_\kappa S^3 + H.c.
\label{eq:Lsoft}
\end{eqnarray}
Specific implementations of the singlet typically involve a subset of these terms.  For example, the 
Next-to-MSSM (NMSSM) \cite{nmssm}
invokes a $Z_3$ symmetry to remove all but the terms involving $\lambda$, $\kappa$, $A_\lambda$, and
$A_\kappa$, whereas the Fat Higgs models  \cite{Harnik:2003rs} have a dynamically generated superpotential
utilizing the $v_0^2$ term to drive electroweak symmetry-breaking, even when supersymmetry is unbroken.
In this work, we do not wed ourselves to any one of these specific realizations, but find that our conclusions can hold for regions of parameter space in any of them.

The additional singlet results in an extra neutralino, and two scalars (one CP even and the other odd) 
in the spectrum.  The parameter space of interest here are regions in which the lightest neutralino is 
largely singlino and the lightest CP even and odd scalars are largely singlets.  In this region, the singlet 
constitutes a kind of hidden sector \cite{hidden} which mixes with the MSSM Higgses through electroweak
symmetry-breaking.  A mostly singlino neutralino is relatively simple to arrange in the limit in which
$( \mu_S + \kappa \langle S \rangle ) \ll M_1, M_2, ( \mu + \lambda \langle S \rangle)$.  A light and mostly
singlet pseudoscalar arises naturally when 
$A_\lambda$, $A_\kappa$, $m_S^2$, $B_S$, $v_0^2$ and $\mu_S$ are small, because
in this limit it is the the pseudo-goldstone boson of  an explicitly broken $U(1)$ symmetry.  
As we will see below, it is also
preferable to have a light and mostly singlet scalar boson.  Unlike the singlino and pseudoscalar, this requires
some engineering of parameters, but the tunings involved are relatively modest, at the $10-20\%$ level.

In the limit of a singlino-like lightest neutralino together with light singlet-like scalar ($h$) and 
pseudoscalar ($a$) Higgs bosons, neutralino annihilations proceeds dominantly to a $ah$ final state 
through $t/u$-channel neutralino exchange and $s$-channel $a$ exchange, with a low velocity cross section given (in the limit of $m_{\chi^0} \gg m_a, m_h$) by:
%
\begin{eqnarray}
\sigma (\chi^0 \chi^0 \rightarrow ah) v
&\approx& 
\frac{1}{64 \pi m^2_{\chi^0}}  \times
\\ &~&
 \hspace*{-3cm}
\bigg[\frac{1}{16 m^2_{\chi^0}} g^2_{haa}  T^2_{a\chi\chi}
+ T^2_{h\chi\chi} T^2_{a\chi\chi}
-\frac{1}{2 m_{\chi^0}}g_{haa}T_{h\chi\chi}T^2_{a\chi\chi} \bigg]
 \nonumber
\label{ha}
\end{eqnarray}
%
%
where $T_{h\chi\chi}$ and $T_{a\chi\chi}$ are the Higgs couplings to the neutralino, $g_{haa}$ is the 
coupling between the Higgs bosons, and $v$ is the relative velocity between the WIMPs. 
In the limit of singlet-like Higgs bosons and a singlino-like neutralino, these couplings reduce to 
$T_{a\chi\chi} \approx T_{h\chi\chi} \approx - \sqrt{2} \kappa$, 
and $g_{haa} \approx \kappa (3 A_{\kappa} -  \mu_S - \kappa \langle S \rangle)$. 
Singlinos can also annihilate to $hh$ or $aa$ final states, but
the rates for these processes are suppressed by $v^2$.
In the regime $g_{haa} / m_{\chi^0} \lsim 1$, the cross section depends only on
$\kappa$ and $m_{\chi^0}$, and
results in a thermal relic abundance of approximately,
\begin{equation}
\Omega_{\chi^0} h^2 \sim 0.1 \times 
\bigg(\frac{0.5}{\kappa}\bigg)^4 
\bigg(\frac{m_{\chi^0}}{200 \,{\rm GeV}}\bigg)^2.
\label{eq:relic}
\end{equation}

If $m_h > 2 m_a$, the scalar Higgs bosons produced in the annihilations will decay dominantly to a pair of 
the pseduoscalar Higgs bosons, leading to a $3 a$ final state. The pseudoscalar Higgs bosons decay via a 
small mixing angle with MSSM Higgs bosons, and may be somewhat long lived. Typically, 
these decays proceed to the heaviest kinematically available fermions. As we are interested in scenarios 
in which neutralino annihilations yield mostly charged leptons, we focus on the case $m_a \lsim$~GeV, 
which leads to the production of either muon pairs ($2 m_{\mu} < m_a < 2 m_{\pi}$) or electron-positron 
pairs ($2 m_{\mu} > m_a$).

To calculate the spectrum of positrons and electrons in the cosmic ray spectrum, we model the diffusion and energy losses of such particles. This is done by solving the steady-state propagation equation~\cite{prop}:
\begin{eqnarray}
0 = \vec{\bigtriangledown} \cdot \bigg[K(E_e)  \vec{\bigtriangledown} \frac{dn_e}{dE_e} \bigg]
+ \frac{\partial}{\partial E_e} \bigg[b(E_e)\frac{dn_e}{dE_e}  \bigg] + Q(E_e,\vec{x}),
\label{dif}
\end{eqnarray}
where $dn_e/dE_e$ is the number density of electrons/positrons per unit energy, $K(E_e)$ is the diffusion coefficient, and $b(E_e,\vec{x})$ is the energy loss rate. The source term, $Q(E_e, \vec{x})$, reflects the mass, annihilation cross section, dominant annihilation modes, and distribution of dark matter in the Galaxy. We adopt a diffusion coefficient of $K(E_e)=6.04\times 10^{28} (E_e/4\,{\rm GeV})^{0.41}$ cm$^2$/s, and boundary conditions corresponding to a disk of 5 kiloparsecs half-thickness, which yields the best-fit to the current body of cosmic ray data (stable and unstable primary-to-secondary nuclei ratios)~\cite{propfit} and an energy loss rate of $b(E_e)=10^{-16} (E_e/{1\,\rm GeV})^2 \, {\rm GeV}$/s, resulting from inverse Compton and synchrotron processes. For the source term, we adopt a Navarro-Frenk-White halo profile to describe the dark matter distribution in the Milky Way. To convert the positron/electron spectrum to a positron fraction, we use the primary and secondary cosmic ray spectra as described in Ref.~\cite{secbg}.

\begin{figure}
\resizebox{9.0cm}{!}{\includegraphics{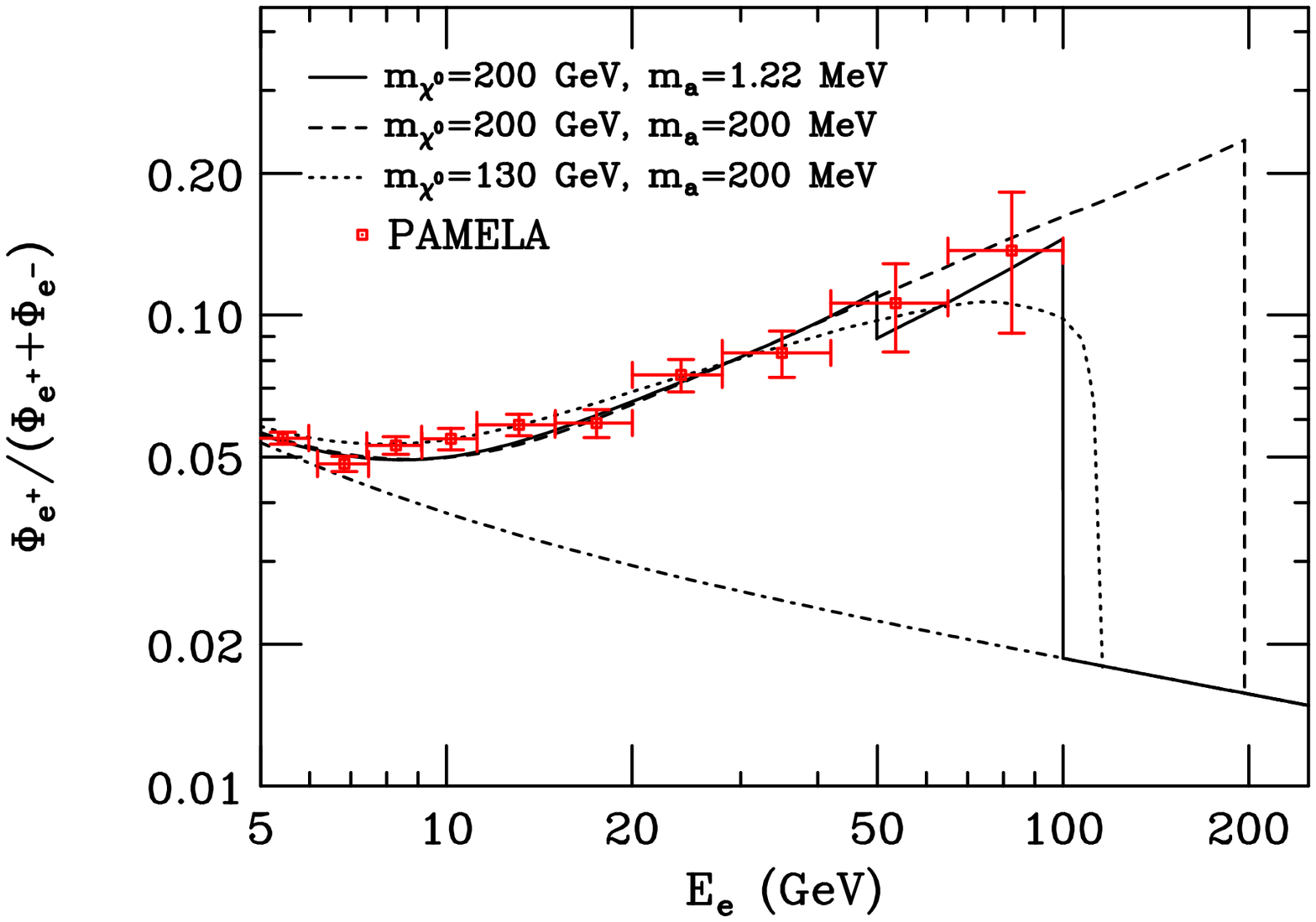}} \\
\resizebox{9.0cm}{!}{\includegraphics{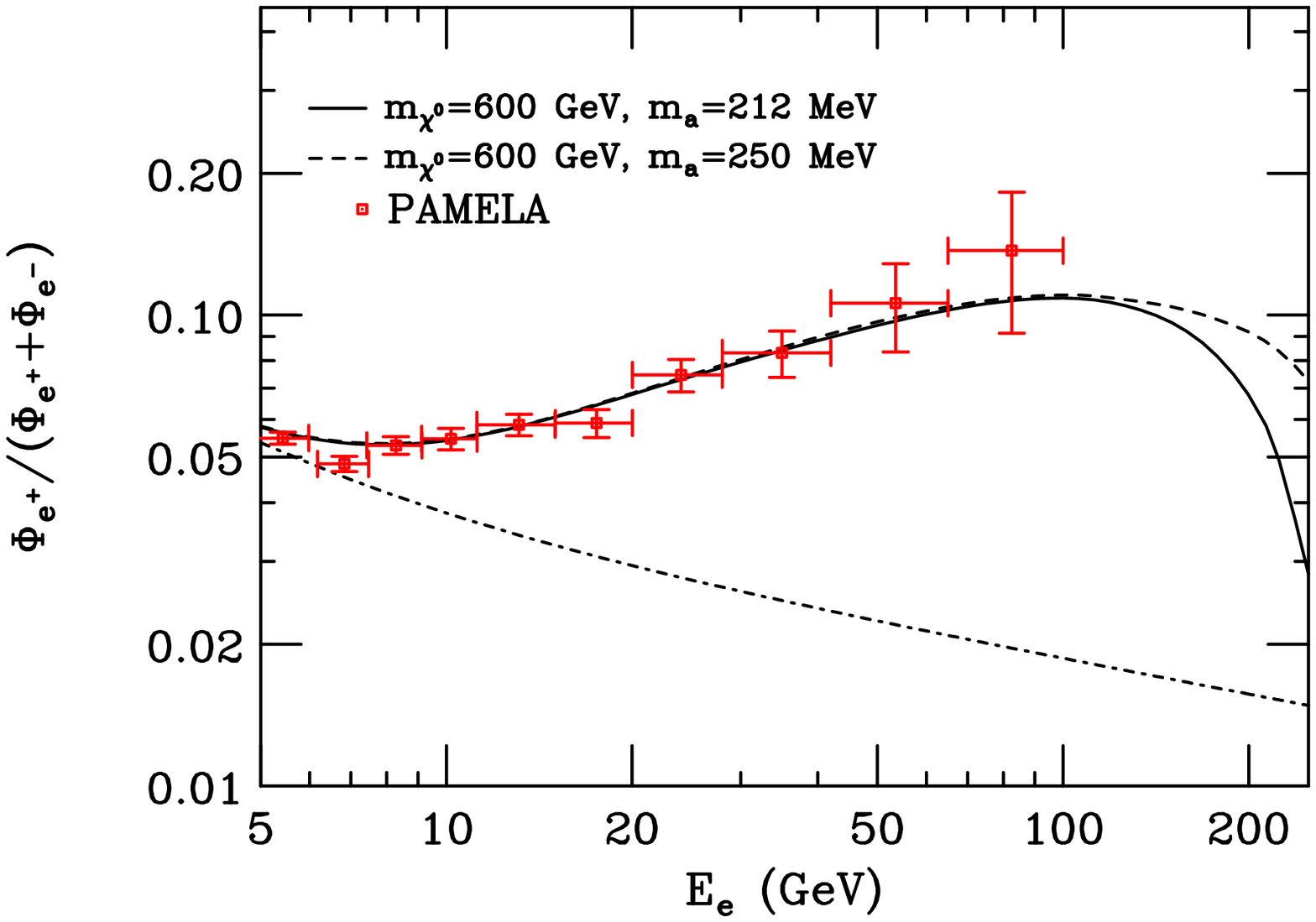}}
\caption{The cosmic ray positron fraction resulting from neutralino annihilations in several selected scenarios. In each case, we consider the channel $\chi^0 \chi^0 \rightarrow ah$, followed by $h \rightarrow aa$, and $a\rightarrow e^+ e^-$ (top) or $\mu^+ \mu^-$ (bottom). In the upper (lower) frame, we have used $m_h=3$ GeV (10 GeV), although the precise value of this mass has only a small effect of the shape of the electron/positron spectrum. In each case, we have normalized the annihilation rate to accommodate the PAMELA data. The dot-dashed line denotes the prediction from astrophysical secondary production alone.}
\label{fig1}
\end{figure}

In Fig.~\ref{fig1}, we show the cosmic ray positron fraction resulting from neutralino annihilations for several choices of the neutralino and pseudoscalar Higgs masses. In the top frame, we consider Higgs decays that produce electron-positron pairs ($m_a < 2 m_{\mu}$), whereas in the lower frame, the decays proceed to muon pairs ($m_a > 2 m_{\mu}$). The mass of the scalar Higgs only mildly impacts the resulting spectrum of positrons.

In each case shown, we have normalized the neutralino annihilation rate to accommodate the PAMELA signal. This rate scales with the neutralino annihilation cross section and with the square of the local dark matter density. The annihilation rate may also be enhanced as a result of inhomogeneities in the dark matter distribution ({\it ie.} clumps or other substructures). Using a local dark matter density of $0.4 \,$GeV/cm$^3$ and an annihilation cross section of $\sigma v =  3 \times 10^{-26}$ cm$^3$/s (required to thermally generate the observed abundance of dark matter), we find that the annihilation rate must be enhanced by a factor of $\sim$10-50 for annihilations to electrons or $\sim$100 for annihilations to muons. 

Fortunately, the rate of neutralino annihilations in the Galactic Halo can be naturally enhanced in this scenario through the Sommerfeld effect generated by the light scalar Higgs boson. Although the resonance structure 
of the Sommerfeld effect can be complex, very roughly speaking, if the light Higgs scalar is lighter than 
$m_h \lsim \kappa^2 m_{\chi^0}/4 \pi$, then for a velocity dispersion expected for the Galactic Halo 
($\sigma \approx 150$\, km/s), the annihilation rate will be enhanced by a factor of about 
$S \sim \kappa^2 m_{\chi^0} / 4 \pi m_h$. More details can be found {\em i.e.}~in the appendices
of Ref.~\cite{sommerfeld2}.

\begin{table}[t]
\begin{tabular}{cc|c|c}
\hline
$m_{\chi^0}$  & $m_a$ & Required BF  & Corresponding $m_h$ \\
\hline \hline
130 GeV & 200 MeV & 40 & $\lsim$ 2 GeV  \\
200 GeV & 1.22 MeV & 11 & $\sim$\,1-4 GeV \\
200 GeV & 200 MeV & 45  & $\lsim$ 4 GeV    \\
600 GeV & 212 MeV & 90 &  $\lsim$ 30 GeV   \\
600 GeV & 250 MeV & 100 & $\lsim$ 30 GeV  \\
\hline \hline
\end{tabular}
\caption{The boost factor to the neutralino annihilation rate required to produce the PAMELA positron excess, for several values of the neutralino and singlet pseudoscalar Higgs masses, and the approximate value (or range) of the singlet scalar Higgs mass required to generate the required boost factor through the Sommerfeld effect.}
\label{param-table}
\end{table}

In Table I, we list the enhancement to the neutralino annihilation rate (relative to that obtained 
for $\sigma v =3 \times 10^{-26}$ cm$^3$/s, no significant substructure, and no Sommerfeld effect) 
required to normalize the positron fraction to the PAMELA observations. For each parameter
set, we determine (assuming $\kappa$ is such that the correct thermal relic density is predicted,
as per Eq.~\ref{eq:relic}) the approximate 
value of the scalar Higgs mass that would lead to a Sommerfeld enhancement providing the required boost 
factor to explain the PAMELA data. 
We find that for $m_{\chi^0} \sim 200$ GeV (600 GeV), we must require $m_h \lsim 4$ 
GeV ($\lsim 30$ GeV) in order to generate the desired annihilation rate.

Light scalars and pseudoscalars must be mostly singlets in order to have escaped detection by previous experiments.  A 30 GeV scalar should be more than about $90\%$ singlet, or it would have been
discovered by LEP II in its Standard Model Higgs search \cite{Barate:2003sz}.  
Tevatron data provides an important constraint through production of a scalar Higgs which decays 
through two light pseudoscalars into a four muon final state \cite{Abazov:2008zz}.  
The null result of this search will be satisfied
for the lightest CP even Higgs provided it has already escaped the LEP bound, and further requires the 
heavier CP even MSSM Higgs bosons to have masses greater than about 135~GeV to suppress the decay into 
pseudoscalars.
For the very light $a$ masses considered here, there is also the possibility of observing rare decays such as
$\Upsilon (3s) \rightarrow \gamma a \rightarrow \gamma \mu^+ \mu^-$ \cite{Dermisek:2006py,Aubert:2009cp}
and $K^+ \rightarrow \pi^+ a \rightarrow \pi^+ \mu^+ \mu^-$ \cite{Bardeen:1978nq,Park:2001cv}.
Both measurements are consistent with an $a$ which is at least $90\%$ singlet, provided $\tan \beta$
is of order one.  

Thus far, we have not addressed the electron (plus positron) spectrum as measured by ATIC~\cite{atic}, and 
more recently by the Fermi Gamma Ray Space Telescope (FGST)~\cite{fgst}. The sharp, edge-like feature 
at $\sim$600 GeV reported by ATIC could easily be accommodated in the scenario discussed here 
for $m_{\chi} \sim 600$ GeV, $m_a < 2 m_{\mu}$, and $m_{h} \lsim 30$ GeV. The spectrum newly 
reported by FGST could also potentially be accommodated, but would require a multi-TeV mass for the 
lightest neutralino~\cite{Bergstrom:2009fa}.

In summary, we have presented a scenario within the context of the Minimal Supersymmetric Standard Model 
extended by a gauge singlet in which the lightest neutralino annihilates to light singlet-like Higgs bosons which
proceed to decay to either electron-positron or muon-antimuon pairs, leading to a cosmic ray positron fraction
consistent with observations of the PAMELA experiment. Furthermore, the annihilation rate of the neutralinos 
in the Galactic Halo can be strongly enhanced by the Sommerfeld effect in this model. No astrophysical boost 
factors are required to obtain the positron fraction observed by PAMELA. 

\medskip

We would like to thank Kathryn Zurek, Bogdan Dobrescu, Paddy Fox, and Bob McElrath for helpful discussions. This work has been supported by the US Department of Energy, including grants DE-FG02-95ER40896
and  DE-AC02-06CH11357, and by NASA grant NAG5-10842.  T~Tait is grateful to the 
SLAC theory group for his many visits

\medskip
{\it Note: As this letter was being finalized, a related study appeared~\cite{Bai:2009ka} which discusses NMSSM neutralinos as a source of the PAMELA excess,  but in a considerably different region of parameter space.}

\end{document}